\documentclass[final,sort&compress]{aipproc}

\layoutstyle{6x9}

\usepackage{subfigure}
\usepackage{natbib}
\bibpunct{(}{)}{;}{a}{}{,}

\begin{document}
\title{Tidal disruption of small satellites orbiting black holes}
\author{C. German\` a}{
  address={Department of Astronomy, University of Padua, Vicolo dell'Osservatorio 2, 35122 Padova, Italy}
  ,altaddress={INAF - Astronomical Observatory of Padova, Vicolo dell'Osservatorio 5, 35122 Padova, Italy},
}
\author{U. Kosti\' c}{
  address={Faculty of Mathematics and Physics, University of Ljubljana, Jadranska 19, 1000 Ljubljana, Slovenia}
}
\author{A. \v Cade\v z}{
  address={Faculty of Mathematics and Physics, University of Ljubljana, Jadranska 19, 1000 Ljubljana, Slovenia}
}
\author{M. Calvani}{
  address={INAF - Astronomical Observatory of Padova, Vicolo dell'Osservatorio 5, 35122 Padova, Italy}
}
\begin{abstract}
Low Mass X-ray Binaries (LMXBs) with either a black hole or a neutron star show power spectra characterised by Quasi Periodic Oscillations (QPOs). Twin peak high frequency QPOs are characterised by frequencies that are typical for matter orbiting within $10\ r_{\mathrm{g}}$ from the compact object. 
We consider clumps of material orbiting a Schwarzschild black hole, that are deformed by tidal interaction. We present some preliminary calculations of corresponding light curves and power spectra. We were able to fit the simulated power spectra with the high frequency part of the power spectra observed in the LMXB XTE J1550-564 containing a black hole. Our numerical simulations reproduce the twin high frequency QPOs and the power-law. The lower peak corresponds to the Keplerian frequency, the upper one to the sum of the Keplerian and the radial frequency.
\end{abstract}
\maketitle
%
%
%
\subsection{QUASI-PERIODIC OSCILLATIONS OBSERVED IN LMXBs}
Power spectra observed in LMXBs, with either a neutron star or a black hole, are characterised by an overall power law shape with superimposed several Lorentzian peaks at different frequencies \citep{2004astro.ph.10551V}. Twin peak high frequency QPOs have central frequencies typical of the Keplerian motion for matter orbiting the compact object within $10\ r_{\mathrm{g}}$ ($r_{\mathrm{g}}$ is the gravitational radius). It is still unclear why the change of source's state is accompanied by the change of frequency, coherence, and amplitude of the peaks \citep{1998ApJ...497L..93K,2006MNRAS.370.1140B}.
Moreover, the central frequencies of this couple of peaks are characterised by the $\sim 3:2$ ratio \citep{2001A&A...374L..19A,2005A&A...437..209B,2008AcA....58...15T,2008NewAR..51..835B}. Low frequency QPOs are observed as well, whose features are thought to be strictly linked with twin high frequency QPOs \citep{1998ApJ...506L..39F,2002ApJ...580.1030R}. First models of these phenomena were proposed by \citet{1998ApJ...508..791M} and \citet{1999ApJ...524L..63S}.
\begin{figure}
\label{fig1}
  \includegraphics[width=\textwidth]{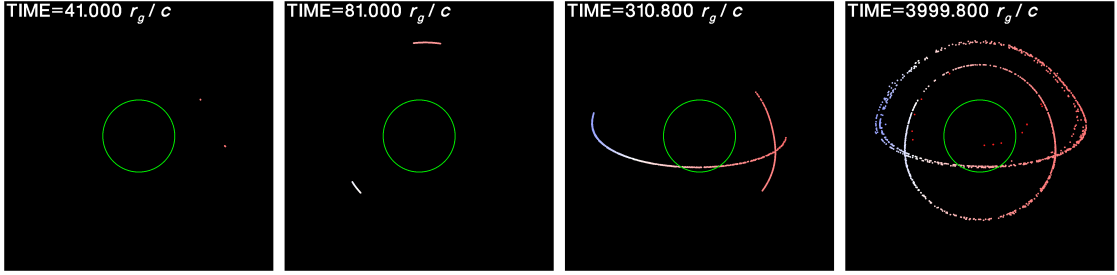}
  \caption{The evolution of tidal deformations of a low-mass satellite orbiting a Schwarzschild black hole as seen by an observer $20^{\circ}$ above the orbital plane. Due to strong gravitational lensing, two images of the satellite can be seen. The colours correspond to the redshift.}
\end{figure}
\section{Tidal interaction of a low-mass satellite with a Schwarzschild black hole}
Tidal evolution of the orbits of low-mass satellites around a Schwarzschild black hole has recently been investigated by \citet{2008A&A...487..527C}. The numerical code we use \citep{2005PhRvD..72j4024C,urosPhDThesis,2009arXiv0901.3447K}, allows to study the effects of the strong gravitational field of a black hole on small objects. The appearance of a spherical blob during its tidal evolution, as calculated by numerical simulation, is shown in Figure~\ref{fig1}. The blob of size of $\sim 0.01\ r_{\mathrm{g}}$ (e.g. $\sim 150\ \mathrm{m}$ for a $10\ M_{\odot}$ black hole) is orbiting a Schwarzschild black hole on an orbit with eccentricity of $e\sim 0.05$ and periastron of $r_p\sim 6.1\ r_{\mathrm{g}}$. The observer is located $20^{\circ}$ above the orbital plane, similar to the inclination of the source XTE J1550-564 constrained by \citet{2002ApJ...568..845O}. Both, the blob and its second-order image due to gravitational lensing, are shown. The blob is squeezed and stretched by tidal forces into a ring-like shape along the orbit. The images are coloured according to the blue/redshift.

A blob orbiting the black hole on a quasi-circular orbit makes radial oscillations along its orbit. \citet{2005ApJ...625..278G} estimated that the energy release during resonant oscillations can be as high as $\sim 0.1\ mc^2 $. Hence, a $\sim 150\ \mathrm{m}$ blob of matter orbiting a $10\ M_{\odot}$ black hole in the inner part of the accretion disk can radiate energy of $\sim 10 ^{34}\ \mathrm{erg}$.
\subsection{Simulated light curve and power spectrum of a tidally disrupted clump of matter}
\begin{figure}\label{fig2}
  \includegraphics[height=0.345\textwidth]{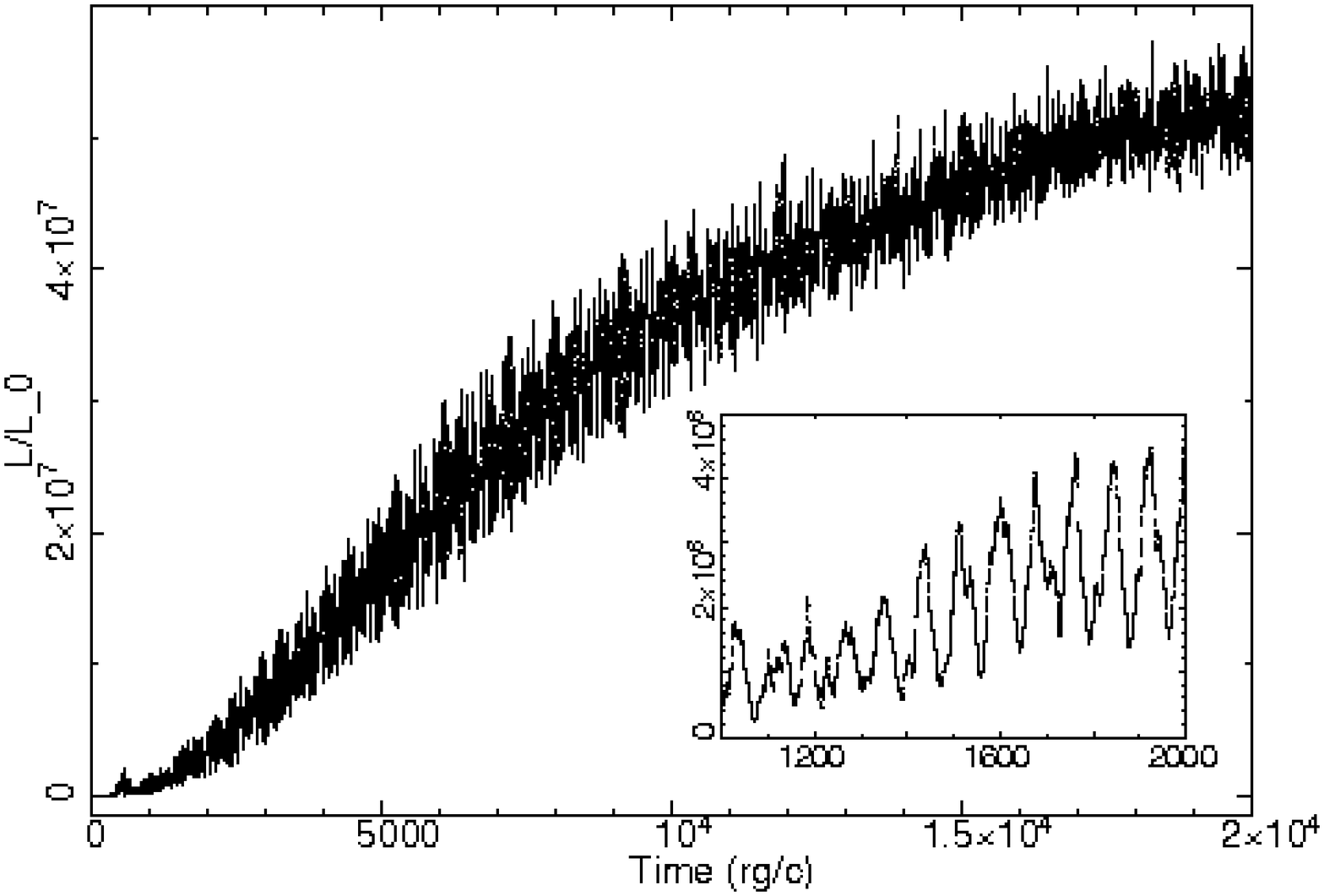}
  \hspace{0.02\textwidth}
  \includegraphics[height=0.345\textwidth]{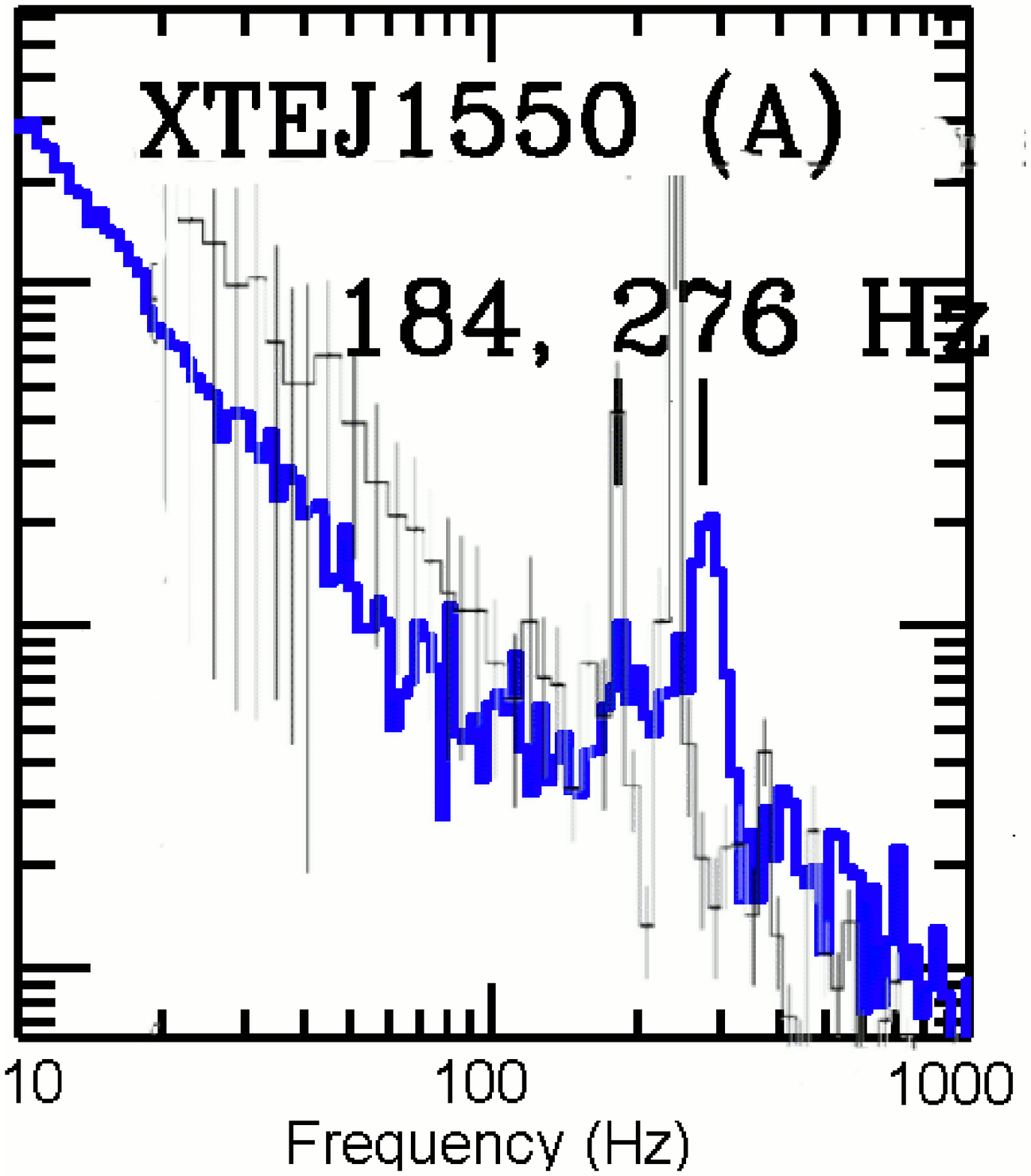}
  \caption{\emph{Left}: simulated light curve produced by a clump of matter orbiting a Schwarzschild black hole. \emph{Right}: Fit of the simulated power spectrum (thin line) and the high frequency part of the observed one of the X-ray binary XTE J1550-564 (thick line) \citep{2006ARA&A..44...49R}.}
\end{figure}

Figure~\ref{fig2} (left) shows the light curve produced by the previous blob. In this region of space-time, the Keplerian frequency ($\nu_k$) is almost three times the radial one ($\nu_r$). During tidal evolution, the light curve shows an overall increase in luminosity which is a consequence of increasing emitting area of the source. The luminosity is normalised to its initial value $L_0$. The time is expressed in $r_{\mathrm{g}}/c$ units. For a $10\ M_{\odot}$ black hole, the light curve lasts for $\sim 1\ \mathrm{s}$. The inset in the figure is an enlargement of a small part of the light curve: each peak takes place on a time scale which is almost the expected Keplerian period at this radius. The peaks are due to relativistic effects, such as gravitational lensing, Doppler boosting and blue/redshift.

Figure~\ref{fig2} (right) shows the fit of the power spectrum of the simulated light curve and the high frequency part of the one observed in LMXB XTE J1550-564 \citep{2006ARA&A..44...49R}. Both twin peak QPOs are reproduced by the model. The lower peak corresponds to $\nu_k$ and the upper one to $\nu_k + \nu_r$. The simulation shows that, at this radial coordinate, the peaks are in a $(\nu_k + \nu_r ) / \nu_k = 1.26$ ratio. Our model reproduces the observed power law without any further assumption. From this fit, we deduce a Schwarzschild black hole mass of $\sim 11\ M_{\odot}$, in agreement with \citet{2002ApJ...568..845O}.

We expect that a better agreement with observations could be obtained by considering the case of a Kerr black hole, however the general results would not change dramatically.
\\
\\
{\footnotesize \emph{Acknowledgements.} We acknowledge support from the bilateral protocol of scientific and technological cooperation between Italy and Slovenia.}
\bibliographystyle{aipproc}
\bibliography{biblio}
\end{document}